\documentclass{ptptex}
\usepackage{wrapft}
\usepackage{graphicx}

\newcommand{\kket}[1]{| {#1} \rangle\!\rangle}     
\newcommand{\wtilde}[1]{\widetilde{#1}} 

\def\bsub{\begin{subequations}}
\def\esub{\end{subequations}}
\def\beq{\begin{eqnarray}}
\def\eeq{\end{eqnarray}}
\def\bsub{\begin{subequations}}
\def\esub{\end{subequations}}
\def\b{\begin{equation}}
\def\bs{\begin{split}}
\def\es{\end{split}}
\def\e{\end{equation}}

\begin{document}

\title{Spontaneous magnetization under a pseudovector interaction between quarks
in high density quark matter}

\author{Masatoshi {\sc Morimoto}$^1$, Yasuhiko {\sc Tsue}$^{2,3}$, {Jo\~ao da {\sc Provid\^encia}}$^{3}$,\\
{Constan\c{c}a {\sc Provid\^encia}}$^{3}$ 
and {Masatoshi {\sc Yamamura}}$^{2,4}$
}

\inst{$^{1}${Graduate School of Integrated Arts and Science, Kochi University, Kochi 780-8520, Japan}\\
$^{2}${Department of Mathematics and Physics, Kochi University, Kochi 780-8520, Japan}\\
$^{3}${CFisUC, Departamento de F\'{i}sica, Universidade de Coimbra, 3004-516 Coimbra, Portugal}\\
$^{4}${Department of Pure and Applied Physics, 
Faculty of Engineering Science,\\
Kansai University, Suita 564-8680, Japan}\\
}

\abst{
Spontaneous magnetization and magnetic susceptibility originated from the pseudovector-type four-point interaction between quarks 
are calculated in quark matter with zero temperature and finite quark chemical potential by using the two-flavor Nambu-Jona-Lasinio model. 
It is shown that both the chiral condensate and spin polarized condensate coexist in a narrow region of the quark chemical potential. 
And then, it is also shown that, in this narrow region, the spontaneous magnetization appears. 
Also, the magnetic susceptibility due to quarks with the positive energy is evaluated in the spin polarized phase. 
}


\maketitle

\section{Introduction}

One of recent interests to understand the world governed by the quantum chromodynamics (QCD) may be to clarify 
the phase structure in the plane spanned by the temperature and baryon chemical potential \cite{FH}. 
In the region of high temperature and zero density, the numerical simulation by using the lattice QCD gives a useful 
information about the phase structure. 
However, in the region with low temperature and large quark chemical potential, 
the lattice simulation does not work until now. 
In that region, it has been remarked that various phases 
may appear such as the color superconducting phase \cite{ARW,IB,CFL}, 
the quarkyonic phase \cite{McL}, the inhomogeneous chiral condensed phase \cite{NT}, 
the quark ferromagnetic phase \cite{Tatsumi}, the color-ferromagnetic phase \cite{Iwazaki}, 
the spin polarized phase due to the axial vector interaction \cite{NMT,TMN} 
or due to the tensor interaction \cite{BJ,IJMP,oursPTP,oursPTEP1,oursPTEP2,oursPTEP3,oursPTEP4,oursPR} 
and so forth. 

Furthermore, it may be interesting to investigate magnetic properties in quark matter 
in the region of low temperature and large baryon chemical potential. 
The reason is as follows: 
In the ultrarelativistic heavy-ion collisions, it has been remarked 
that a strong magnetic field may be created in the early stage of nucleus-nucleus collisions \cite{KMW}, 
for example, $|eB|\sim m_{\pi}^2$ at the Relativistic Heavy-Ion Collider (RHIC) experiment at Brookhaven, 
where $e$, $B$ and $m_{\pi}$ represent the elementary electric charge, the magnetic 
flux density or magnetic field and pion mass, respectively, and maybe even stronger at the Large Hadron Collider (LHC) experiment 
at CERN. 
In astrophysical fields, compact stars such as neutron stars, especially magnetars \cite{magnetar1,magnetar2}, 
show a very strong magnetic field.  
Thus, the investigation of magnetic properties in quark matter is one of the interesting and important problems of QCD 
\cite{Review1,Review2}. 

Recently, the present authors have investigated the phase structure of high density quark matter under a strong external magnetic field \cite{Ours5}.
By using the Nambu-Jona-Lasinio (NJL) model with tensor or pseudovector interaction between quarks, 
we have shown that a quark spin polarized phase may exist in high density quark matter under a strong external magnetic field 
in a certain model parameter regions. 
Similarly, the effects of a strong magnetic field in the model with the tensor interaction have been investigated 
in one-flavor NJL model at finite temperature with zero chemical potential \cite{Ferrer,Ferrer2}. 
Also, in Ref. \citen{fukushima}, the effects of the axial-vector interaction under a strong magnetic field on the spatially-modulated chiral condensed phase 
was investigated by means of the holographic technique. 
Thus, the physics of the strong interacting matter under a magnetic field becomes interesting and important subject in quark matter with various 
possible phases and many investigations are carried out recently \cite{text}.

In this paper, we investigate spontaneous magnetization in the finite quark chemical potential region or in high density quark matter at zero temperature 
by using the Nambu-Jona-Lasinio (NJL) model \cite{NJL,Buballa,Klevansky,HK} 
with the pseudovector-type \cite{NMT} four-point interaction between quarks  
as an effective model of QCD. 
As for the tensor-type interaction, we have already investigated a possibility of a spontaneous magnetization \cite{oursPTEP3}. 
As a result, the tensor interaction does not reveal the spontaneous magnetization except for the existence of the 
anomalous magnetic moments of quarks, even if the spin polarized condensate exists. 
Thus, in this paper, we investigate the magnetic properties due to the pseudovector interaction between quarks in quark matter at zero temperature.

This paper is organized as follows: 
In the next section, a model under consideration is introduced as an extension of the original NJL model.  
In Sect. 3, the thermodynamic potential is evaluated under a weak external magnetic filed and 
the way how to calculate the spontaneous magnetization is explained. 
In Sect. 4, numerical results are shown for the chiral condensate, spin polarized condensate and the thermodynamic potential, 
and the spontaneous magnetization and the magnetic susceptibility are calculated. 
The last section is devoted to a summary and concluding remarks.

\section{Mean field approximation for the Nambu-Jona-Lasinio model with vector-pseudovector-type four-point interactions between quarks}

Let us start from the two-flavor Nambu-Jona-Lasinio model with vector-pseudovector-type \cite{NMT,TMN} 
four-point interactions between quarks under an external magnetic field. 
The Lagrangian density can be expressed as 
\beq\label{2-1}
{\cal L}&=&
{\bar \psi}(i\gamma^\mu D_{\mu}-m_0)\psi+G_s[({\bar \psi}\psi)^2+({\bar \psi}i\gamma_5{\vec \tau}\psi)^2]
\nonumber\\
& &
-G_p[({\bar \psi}\gamma^\mu{\vec \tau}\psi)^2+({\bar \psi}i\gamma_5\gamma^\mu{\vec \tau}\psi)^2]\ ,
\eeq
where $m_0$ represents a current quark mass and $D_\mu$ represents the covariant derivative introduced as 
\beq\label{2-2}
D_\mu=\partial_\mu+iQA_\mu\ ,\qquad 
A_\mu=\left(0,\ \frac{By}{2},\ -\frac{Bx}{2},\ 0\right) = (0,\ -{\mib A})\  .
\eeq
Here, $Q=2e/3$ for up quark and $-e/3$ for down quark are the electric charges where $e$ is the elementary charge.
There is an external magnetic field $B$ along $z$-axis.  

Hereafter, we treat the model within the mean field approximation. 
In order to consider the spin polarization under the mean field approximation, 
the pseudovector condensate $\langle {\bar \psi}\gamma_5\gamma^3\tau_3\psi\rangle$
is taken into account. 
Then, the Lagrangian density reduces to 
\beq\label{2-3}
{\cal L}_{MF}&=&{\bar \psi}(i\gamma^\mu D_\mu -M_q)\psi+U_A{\bar \psi}\gamma_5\gamma^3\tau_3\psi
-\frac{M^2}{4G_s}-\frac{U_A^2}{4G_p}
\nonumber\\
&=&{\bar \psi}(i\gamma^\mu D_\mu -M_q)\psi-U_A{\psi^{\dagger}}\Sigma_3\tau_3\psi
-\frac{M^2}{4G_s}-\frac{U^2}{4G_p} \ , 
\eeq
where 
\beq\label{2-4}
& &\Sigma_3=-\gamma^0\gamma_5\gamma^3=
\left(
\begin{array}{cc}
\sigma_3 & 0 \\
0 & \sigma_3
\end{array}
\right)
\nonumber\\
& &M_q=m_0+M\ , \qquad M=-2G_s\langle{\bar \psi}\psi\rangle\ , \nonumber\\
& &U_A=2G_p\langle {\bar \psi}\gamma_5\gamma^3\tau_3\psi\rangle=-2G_p\langle\psi^{\dagger}\Sigma_3\tau_3\psi\rangle\equiv U\tau_f\ . 
\eeq
Here, $\tau_f=1$ for up quark and $-1$ for down quark denote the eigenvalues of $\tau_3$. 
Also, $\sigma_3$ is the third component of the Pauli spin matrices.

Introducing the quark chemical potential $\mu$ in order to consider finite density quark matter, 
the Hamiltonian density can be obtained from the Lagrangian density within the mean field approximation as 
\beq\label{2-5}
{\cal H}_{MF}-\mu{\cal N}
&=&{\bar \psi}\left(-i{\mib \gamma}\cdot ({\mib \nabla}-iQ{\mib A})+M_q-\mu\gamma^0+U_A\gamma^0\Sigma_3\tau_3
\right)\psi
\nonumber\\
& &
+\frac{M^2}{4G_s}+\frac{U^2}{4G_p}\ ,
\eeq
where ${\cal N}$ represents the quark number density, $\psi^{\dagger}\psi$.

\section{Thermodynamic potential}

The Hamiltonian density (\ref{2-5}) 
can be rewritten as 
\beq\label{3-1}
{\cal H}_{MF,A}-\mu{\cal N}
&=&\psi^{\dagger}(h_A-\mu)\psi +\frac{M^2}{4G_s}+\frac{U^2}{4G_p}\ ,\\
h_A&=&-i\gamma^0{\mib \gamma}\cdot({\mib \nabla}-iQ{\mib A})+\gamma^0 M_q+U_A\Sigma_3\tau_3\ . 
\eeq
In order to obtain the eigenvalues of $h_A$, namely the energy eigenvalues of a single quark, it is necessary to diagonalize $h_A$, 
the eigenvalues of which can be obtained easily as 
\beq\label{3-2}
E_{A,p\nu\eta}^{f}&=&
\left\{
\begin{array}{l}
{\displaystyle 
E_{A,p\nu\sigma}^u=\sqrt{2Q_u B\nu+\left(\sqrt{p_z^2+M_q^2}+\sigma U\right)^2}\ , 
\left\{
\begin{array}{ll}
\nu=0,1,2,\cdots\ & {\rm for}\ \ \sigma=1 \\
\nu=1,2,\cdots \ & {\rm for}\ \ \sigma=-1
\end{array}
\right.}\\
{\displaystyle
E_{A,p\nu\sigma}^d=\sqrt{-2Q_d B\nu+\left(\sqrt{p_z^2+M_q^2}-\sigma U\right)^2}\ , 
\left\{
\begin{array}{ll}
\nu=1,2,\cdots & {\rm for}\ \ \sigma=1 \\
\nu=0,1,2,\cdots  & {\rm for}\ \ \sigma=-1
\end{array}
\right.}
\end{array}\right.
\nonumber\\
&=&
\sqrt{2|Q_f|B\nu+\left(\eta U+\sqrt{p_z^2+M_q^2}\right)^2}\ .
\quad
\left\{
\begin{array}{ll}
\nu=0, 1,2,\cdots & {\rm for}\ \  \eta=1 \\
\nu=1,2,\cdots & {\rm for}\ \ \eta=-1
\end{array}
\right.
\eeq 
The thermodynamic potential can be expressed as
\beq\label{3-3}
\Phi_A&=&
\sum_{f,\alpha}\int\frac{dp_z}{2\pi}\frac{|Q_f|B}{2\pi}\left(E_{A,p\ \nu=0\ \eta=1}^f-\mu\right)
\theta(\mu-E_{A,p\ \nu=0\ \eta=1}^f)
\nonumber\\
& &
+
\sum_{\eta,f,\alpha}\int\frac{dp_z}{2\pi}\sum_{\nu=1}^{E_A<\mu}\frac{|Q_f|B}{2\pi}\left(E_{A,p\nu\eta}^f-\mu\right)
\theta(\mu-E_{A,p\nu\eta}^f)
\nonumber\\
& &-\sum_{f,\alpha}\int^{\Lambda}\frac{dp_z}{2\pi}\frac{|Q_f|B}{2\pi}E_{A,p\ \nu=0\ \eta=1}^f
-
\sum_{\eta,f,\alpha}\int^{\Lambda}\frac{dp_z}{2\pi}\sum_{\nu=1}^{E_A<\Lambda}\frac{|Q_f|B}{2\pi}E_{A,p\nu\eta}^f
\nonumber\\
& &
+\frac{M^2}{4G_s}+\frac{U^2}{4G_p}\ . 
\eeq 
The first and the second lines represent the positive-energy contribution of quarks and the third line represents 
the vacuum contribution. 
It should be noted that the single quark energy does not depend on the flavor in the lowest Landau level with $\nu=0$.


In the thermodynamic potential (\ref{3-3}), 
the quantum number $\nu$, which labels the Landau level, has to be summed up.  
However, since it is interesting to consider the spontaneous magnetization, it may be assumed that the external magnetic field $B$ 
is small and finally $B$ becomes 0.
Therefore, let us replace the sum with respect to $\nu$ by an integration approximately \cite{BJ}. 
In general, let us consider a function $f(x)$. 
Here, we introduce a small quantity $a$ and let us consider the Tailor expansion around $x=a\nu$ as follows:
\beq\label{4-1}
\int_{a(\nu-1)}^{a\nu}dx f(x)
&=&
\int_{a(\nu-1)}^{a\nu}dx \left[
f(a\nu)+\left.\frac{df}{dx}\right|_{x=a\nu}(x-a\nu)+\left.\frac{1}{2}\frac{d^2 f}{dx^2}\right|_{x=a\nu}(x-a\nu)^2+\cdots\right]
\nonumber\\
&=&af(a\nu)-\frac{1}{2}a^2f'(a\nu)+\frac{1}{6}a^3f''(a\nu)+\cdots\ .
\eeq 
Thus, the following relations is obtained :
\beq\label{4-2}
\sum_{\nu=\nu_m+1}^{\nu_M}\int_{a(\nu-1)}^{a\nu}dx f(x) &\equiv& \int_{a\nu_m}^{a\nu_M}dx f(x)\nonumber\\
&=&
a\!\!\sum_{\nu=\nu_m+1}^{\nu_M}\!\!f(a\nu)-\frac{a^2}{2}\!\!\sum_{\nu=\nu_m+1}^{\nu_M}\!\!f'(a\nu)+\frac{a^3}{6}\!\!\sum_{\nu=\nu_m+1}^{\nu_M}\!\!f''(a\nu)+\cdots\ . 
\nonumber\\
& & 
\eeq
Here, it should be noted that the definition of integral can be used when $a$ is infinitesimally small, namely, 
\beq\label{4-3}
a\sum_{\nu=\nu_m+1}^{\nu_M}f'(a\nu)&=&\int_{a\nu_m}^{a\nu_M}dx\ f'(x) +\frac{a^2}{2}\sum_{\nu=\nu_m+1}^{\nu_M}f''(a\nu)+\cdots
\nonumber\\
&=&f(a\nu_M)-f(a\nu_m)+\frac{a^2}{2}\sum_{\nu=\nu_m+1}^{\nu_M}f''(a\nu)+\cdots\ , 
\eeq
and so on. 
Thus, by using the above formula repeatedly, useful approximate formula is obtained as follows:
\beq\label{4-4}
& &a\sum_{\nu=\nu_m+1}^{\nu_M}f(a\nu)\nonumber\\
&=&\int_{a\nu_m}^{a\nu_M}dx f(x)+\frac{a}{2}\left[f(a\nu_M)-f(a\nu_m)\right]
+\frac{a^2}{12}\left[f'(a\nu_M)-f'(a\nu_m)\right]+\cdots\ .
\eeq
In (\ref{3-3}), we separate the sum over $\nu$ into a part with $\nu=0$ and another one with $\nu > 0$. 
As for the positive-energy part with $\eta=1$, we obtain 
\beq\label{4-5}
\nu_m=0\ , \qquad \nu_M\equiv \nu_M^{(1)}=\left[\frac{\mu^2-\left(\sqrt{p_z^2+M_q^2}+U\right)^2}{2|Q_f|B}\right]\ , 
\eeq
where $[\cdots ]$ represents the Gauss symbol. 
Also, $\nu_M\geq 0$, we obtain 
\beq\label{4-6}
|p_z| \leq \sqrt{(\mu-U)^2-M_q^2}\ . 
\eeq
Similarly, for $\eta=-1$, we
obtain 
\beq\label{4-7}
& &\nu_m=0\ , \qquad \nu_M\equiv \nu_M^{(-1)}=\left[\frac{\mu^2-\left(\sqrt{p_z^2+M_q^2}-U\right)^2}{2|Q_f|B}\right]\ , \nonumber\\
& &|p_z| \leq \sqrt{(\mu+U)^2-M_q^2}\ .
\eeq
As for the vacuum contributions, the three-momentum cutoff $\Lambda$ is, as usually, introduced as 
\beq\label{4-8}
p_x^2+p_y^2+p_z^2 \leq \Lambda^2\ . 
\eeq
In the case under consideration, the Landau quantization in the $x$-$y$-plane is carried out and 
$p_x^2+p_y^2$ should be replaced to $2|Q_f|B\nu$. 
Thus, we get the maximum integer of $\nu$ as 
\beq\label{4-9}
\nu_M\equiv \nu_M^{\rm vac}=\left[\frac{\Lambda^2-p_z^2}{2|Q_f|B}\right]\ , \qquad |p_z|\leq \Lambda\ . 
\eeq
Thus, the thermodynamic potential (\ref{3-3}) can be evaluated, for example, up to order of $B$ as follows:
\beq\label{4-10}
\Phi&=&
\Phi_0+\Phi_{1}+\Phi_{-1}\ , \\
\Phi_0&=&\sum_{f,\alpha}\int\frac{dp_z}{2\pi}\frac{|Q_f|B}{2\pi}\left(E_{A,p\ \nu=0\ \eta=1}^f-\mu\right)\theta(\mu-E_{A,p\ \nu=0\ \eta=1}^f)\nonumber\\
& &-\sum_{f,\alpha}\int \frac{dp_z}{2\pi}\frac{|Q_f|B}{2\pi}E_{A,p\ \nu=0\ \eta=1}^f
+\frac{M^2}{4G_s}+\frac{U^2}{4G_p}\nonumber\\
&=&\frac{3eB}{4\pi^2}\left[-(\mu-U)\sqrt{(\mu-U)^2-M_q^2}+M_q^2\ln\frac{\mu-U+\sqrt{(\mu-U)^2-M_q^2}}{M}\right]
\nonumber\\
& &
\qquad\qquad\qquad\qquad\qquad\qquad\qquad\qquad\qquad\qquad\qquad\qquad
\times
\theta(\mu-(U+M_q))\nonumber\\
& &-\frac{3eB}{4\pi^2}\left[\Lambda\sqrt{\Lambda^2+M_q^2}+M_q^2\ln\frac{\Lambda+\sqrt{\Lambda^2+M_q^2}}{M}+2U\Lambda\right]
+\frac{M^2}{4G_s}+\frac{U^2}{4G_p}\ , \nonumber
\eeq
\beq
\Phi_1&=&\sum_{f,\alpha}\int\frac{dp_z}{2\pi}\sum_{\nu=1}^{\nu_M^1}\frac{|Q_f|B}{2\pi}\left(E_{A, p\nu\ \eta=1}^f-\mu\right)\theta(\mu-E_{A,p\nu\ \eta=1}^f)
+\Phi_1^{\rm vac}\nonumber\\
&=&\frac{3}{4\pi^2}\Biggl[
\frac{1}{6}\sqrt{(\mu-U)^2-M_q^2}\left(-2\mu^3+2\mu^2U+2\mu U^2-2U^3-13M_q^2 U+5\mu M_q^2\right)\nonumber\\
& &\qquad\qquad
-\frac{M_q^2}{2}\left(M_q^2+4U^2-4\mu U\right)\ln\frac{\mu-U+\sqrt{(\mu-U)^2-M_q^2}}{M_q}\Biggl]
\nonumber\\
& &
\qquad\qquad\qquad\qquad\qquad\qquad\qquad\qquad\qquad\qquad\qquad\qquad
\times
\theta(\mu-(M_q+U))\nonumber\\
& &
+\frac{3eB}{8\pi^2}\left[(\mu-U)\sqrt{(\mu-U)^2-M_q^2}-M_q^2\ln\frac{\mu-U+\sqrt{(\mu-U)^2-M_q^2}}{M_q}\right]
\nonumber\\
& &
\qquad\qquad\qquad\qquad\qquad\qquad\qquad\qquad\qquad\qquad\qquad\qquad
\times\theta(\mu-(M_q+U))\nonumber\\
& &+\Phi_1^{\rm vac}\ , \nonumber
\\
\Phi_1^{\rm vac}&=&-\sum_{f,\alpha}\int \frac{dp_z}{2\pi}\sum_{\nu=1}^{\nu_M^{\rm vac}}\frac{|Q_f|B}{2\pi}E_{A,p\nu\ \eta=1}\nonumber\\
&=&-\frac{1}{\pi^2}\int_0^{\Lambda}dp_z\left[\left(\Lambda^2-p_z^2+\left(\sqrt{p_z^2+M_q^2}+U\right)^2\right)^{\frac{3}{2}}
-\left(\sqrt{p_z^2+M_q^2}+U\right)^3\right]\nonumber\\
& &-\frac{3eB}{4\pi^2}
\int_0^{\Lambda}dp_z\left[\sqrt{\Lambda^2-p_z^2+\left(\sqrt{p_z^2+M_q^2}+U\right)^2}-\left(\sqrt{p_z^2+M_q^2}+U\right)\right]\ , \nonumber\\
\Phi_{-1}&=&\sum_{f,\alpha}\int\frac{dp_z}{2\pi}\sum_{\nu=1}^{\nu_M^{-1}}\frac{|Q_f|B}{2\pi}\left(E_{A, p\nu\ \eta=-1}^f-\mu\right)
\theta(\mu-E_{A,p\nu\ \eta=-1}^f)
+\Phi_{-1}^{\rm vac}\nonumber\\
&=&\frac{3}{4\pi^2}\Biggl[
\frac{1}{6}\sqrt{(\mu+U)^2-M_q^2}\left(-2\mu^3-2\mu^2U+2\mu U^2+2U^3+13M_q^2 U+5\mu M_q^2\right)\nonumber\\
& &\qquad\qquad
-\frac{M_q^2}{2}\left(M_q^2+4U^2+4\mu U\right)\ln\frac{\mu+U+\sqrt{(\mu+U)^2-M_q^2}}{M_q}\Biggl]
\nonumber\\
& &
\qquad\qquad\qquad\qquad\qquad\qquad\qquad\qquad\qquad\qquad\qquad\qquad
\times
\theta(\mu-(M_q-U))\nonumber\\
& &
+\frac{3eB}{8\pi^2}\left[(\mu+U)\sqrt{(\mu+U)^2-M_q^2}-M_q^2\ln\frac{\mu+U+\sqrt{(\mu+U)^2-M_q^2}}{M_q}\right]
\nonumber\\
& &
\qquad\qquad\qquad\qquad\qquad\qquad\qquad\qquad\qquad\qquad\qquad\qquad
\times
\theta(\mu-(M_q-U))\nonumber\\
& &+\Phi_{-1}^{\rm vac}\ , \nonumber
\\
\Phi_{-1}^{\rm vac}&=&-\sum_{f,\alpha}\int \frac{dp_z}{2\pi}\sum_{\nu=1}^{\nu_M^{\rm vac}}\frac{|Q_f|B}{2\pi}E_{A,p\nu\ \eta=-1}\nonumber\\
&=&-\frac{1}{\pi^2}\int_0^{\Lambda}dp_z\left[\left(\Lambda^2-p_z^2+\left(\sqrt{p_z^2+M_q^2}-U\right)^2\right)^{\frac{3}{2}}
-\left(\sqrt{p_z^2+M_q^2}-U\right)^3\right]\nonumber\\
& &-\frac{3eB}{4\pi^2}
\int_0^{\Lambda}dp_z\left[\sqrt{\Lambda^2-p_z^2+\left(\sqrt{p_z^2+M_q^2}-U\right)^2}-\left(\sqrt{p_z^2+M_q^2}-U\right)\right] . 
\eeq 
Here, the integrations $(\sqrt{p_z^2+M_q^2}+U)^n$ in $\Phi_1^{\rm vac}$ and $\Phi_{-1}^{\rm vac}$ can be performed. 
As a result, the thermodynamic potential $\Phi_A$ is arranged with the term independent of $B$ and dependent of $B$:
\beq
\Phi_A&=&\Phi_{B=0}+\Phi_{B}+\Phi_{B^2}+O(B^3)\ , 
\label{21}
\eeq
\beq
\Phi_{B=0}&=&
\frac{3}{4\pi^2}\Biggl[
\frac{1}{6}\sqrt{(\mu-U)^2-M_q^2}\left(-2\mu^3+2\mu^2 U+2\mu U^2-2U^3-13M_q^2 U+5\mu M_q^2\right)\nonumber\\
& &\qquad
-\frac{M_q^2}{2}(M_q^2+4U^2-4\mu U)\ln\frac{\mu-U+\sqrt{(\mu-U)^2-M_q^2}}{M_q}\Biggl]\theta(\mu-(M_q+U))\nonumber\\
& &+
\frac{3}{4\pi^2}\Biggl[
\frac{1}{6}\sqrt{(\mu+U)^2-M_q^2}\left(-2\mu^3-2\mu^2 U+2\mu U^2+2U^3+13M_q^2 U+5\mu M_q^2\right)\nonumber\\
& &\qquad
-\frac{M_q^2}{2}(M_q^2+4U^2+4\mu U)\ln\frac{\mu+U+\sqrt{(\mu+U)^2-M_q^2}}{M_q}\Biggl]\theta(\mu-(M_q-U))
\nonumber\\
& &+\frac{1}{4\pi^2}
\left[\Lambda\sqrt{\Lambda^2+M_q^2}(5M_q^2+2\Lambda^2+12U^2)+3M_q^2(M_q^2+4U^2)\ln\frac{\Lambda+\sqrt{\Lambda^2+M_q^2}}{M_q}\right]
\nonumber\\
& &-\frac{1}{\pi^2}\int_0^{\Lambda}dp_z
\Biggl[\left(\Lambda^2-p_z^2+\left(\sqrt{p_z^2+M_q^2}-U\right)^2\right)^{\frac{3}{2}}
\nonumber\\
& &\qquad\qquad\qquad\qquad
+\left(\Lambda^2-p_z^2+\left(\sqrt{p_z^2+M_q^2}+U\right)^2\right)^{\frac{3}{2}}\Biggl]
+\frac{M^2}{4G_s}+\frac{U^2}{4G_p}\ , 
\label{4-11}
\\
\Phi_B&=&
\frac{3eB}{8\pi^2}\left[-(\mu-U)\sqrt{(\mu-U)^2-M_q^2}+M_q^2\ln\frac{\mu-U+\sqrt{(\mu-U)^2-M_q^2}}{M_q}\right]
\nonumber\\
& &
\qquad\qquad\qquad\qquad\qquad\qquad\qquad\qquad\qquad\qquad\qquad\qquad
\times
\theta(\mu-(M_q+U))\nonumber\\
& &+
\frac{3eB}{8\pi^2}\left[(\mu+U)\sqrt{(\mu+U)^2-M_q^2}-M_q^2\ln\frac{\mu+U+\sqrt{(\mu+U)^2-M_q^2}}{M_q}\right]\nonumber\\
& &
\qquad\qquad\qquad\qquad\qquad\qquad\qquad\qquad\qquad\qquad\qquad\qquad
\times
\theta(\mu-(M_q-U))\nonumber\\
& &-\frac{3eB}{2\pi^2}{\Lambda U}
\nonumber\\
& &
-\frac{3eB}{4\pi^2}\int_0^{\Lambda}dp_z\left[\sqrt{\Lambda^2-p_z^2+\left(\sqrt{p_z^2+M_q^2}+U\right)^2}
+\sqrt{\Lambda^2-p_z^2+\left(\sqrt{p_z^2+M_q^2}-U\right)^2}\right]\ , \nonumber\\
& &\label{22}
%
\eeq
\beq
	\Phi_{B^2} &= &\frac{5e^2B^2}{72\pi^2}\left[\frac{\sqrt{(\mu-U)^2-M_q^2}}{\mu}-\int^{\sqrt{(\mu-U)^2-M_q^2}}_0 dp_z \frac{1}{\sqrt{p_z^2+M_q^2}+U} \right]
\nonumber\\
& &
\qquad\qquad\qquad\qquad\qquad\qquad\qquad\qquad\qquad\qquad\qquad\qquad
\times
 \theta\left(\mu-(M_q+U)\right) \nonumber \\
	& &+\frac{5e^2B^2}{72\pi^2}\left[\frac{\sqrt{(\mu+U)^2-M_q^2}}{\mu}-\int^{\sqrt{(\mu+U)^2-M_q^2}}_0 dp_z \frac{1}{\sqrt{p_z^2+M_q^2}-U} \right]
\nonumber\\
& &
\qquad\qquad\qquad\qquad\qquad\qquad\qquad\qquad\qquad\qquad\qquad\qquad
\times
 \theta\left(\mu-(M_q-U)\right) 
\nonumber
\\
	& &-\frac{5e^2B^2}{72\pi^2}\int^\Lambda_0 dp_z \Biggl[ \left( \Lambda^2 - p_z^2 + \left( \sqrt{p_z^2+M_q^2} + U\right)^2\right)^{-\frac{1}{2}} 
\nonumber\\
& &\qquad\qquad\qquad\qquad
+ \left( \Lambda^2 - p_z^2 + \left( \sqrt{p_z^2+M_q^2} - U\right)^2\right)^{-\frac{1}{2}} \Biggl] \nonumber \\
	& &+\frac{5e^2B^2}{72\pi^2}\int^\Lambda_0 dp_z \left[\frac{1}{\sqrt{p_z^2+M_q^2}+U}+\frac{1}{\sqrt{p_z^2+M_q^2}-U}\right] \ .\nonumber\\
& &\label{4-12}
\eeq
Here, it is found that no problem arises. 
We define a spontaneous magnetization ${\cal M}$ as 
\beq\label{4-13}
{\cal M}=-\left.\frac{\partial \Phi_A}{\partial B}\right|_{B=0}\ . 
\eeq
Here, in (\ref{22}), even if $U=0$ and $\mu <M_q$, 
${\cal M}$ appears because the last two integrations in the last line in (\ref{22}) survives, which 
leads to $\Phi_B=-3eB/(2\pi^2)\cdot \Lambda\sqrt{\Lambda^2+M_q^2}$. 
When we sum up $\nu$ over the Landau level, we have introduced the maximum value of $\nu$.  
Then, the maximum value of $\nu$, namely $\nu_M^{\rm vac}$, has been replaced to the boundary value $(\Lambda^2-p_z^2)/(2|Q_f|B)$ which 
is not always integer. 
So, we subtract the following $\Phi_R$ in order to delete this artificial contribution. 
\beq\label{4-14}
\Phi&=&\Phi_A-\Phi_R\ , 
\\
\Phi_R&=&\sum_{f,\alpha}\int_{-\Lambda}^{\Lambda} \frac{dp_z}{2\pi}\frac{|Q_f|B}{2\pi}(-E_\Lambda)
=-\frac{3eB}{2\pi^2}\Lambda\sqrt{\Lambda^2+M_q^2}\ , 
\nonumber
\eeq
where $E_\Lambda=\sqrt{\Lambda^2+M_q^2}$. 
As a result, ${\cal M}$ disappears when $U=0$ and $\mu<M_q$.

\section{Numerical results}

\subsection{Spontaneous magnetization}

First, we set up $U=0$. Then, $\Phi_{B=0}$ is written as
\beq\label{5-1}
\Phi_{B=0}(U=0)
&=&\frac{3}{4\pi^2}\left[\frac{1}{3}\sqrt{\mu^2-M_q^2}(-2\mu^3+5\mu M_q^2)-M_q^4\ln\frac{\mu+\sqrt{\mu^2-M_q^2}}{M_q}\right]
\nonumber\\
& &
\qquad\qquad\qquad\qquad\qquad\qquad\qquad\qquad\qquad\qquad\qquad\qquad
\times
\theta(\mu-M_q)\nonumber\\
& &-\frac{3}{4\pi^2}\left[\Lambda\sqrt{\Lambda^2+M_q^2}(2\Lambda^2+M_q^2)-M_q^4\ln\frac{\Lambda+\sqrt{\Lambda^2+M_q^2}}{M_q}\right]
\nonumber\\
& &
+\frac{M^2}{4G_s}\ . 
\eeq 
When we adopt the chiral limit, namely $M_q=M$ with $m_0=0$, the gap equation is derived as 
\beq\label{5-2}
\frac{\partial \Phi_{B=0}(U=0)}{\partial M}
&=&
-\frac{3M}{\pi^2}\left[\Lambda\sqrt{\Lambda^2+M^2}-M^2\ln\frac{\Lambda+\sqrt{\Lambda^2+M^2}}{M}-\frac{\pi^2}{6G_s}\right]
\nonumber\\
&=&0 \ .
\eeq
There is a solution except for $M=0$ in the vacuum $\mu=0$. 
For example, if we adopt the model parameters 
$\Lambda=0.631$ GeV and $G_s=5.5$ GeV$^{-2}$, then, 
the dynamical quark mass $M$ is obtained as $M=0.322$ GeV. 
If we introduce the current quark mass $m_0=0.005$ GeV, the constituent quark mass $M_q=0.335$ GeV is obtained  
under the same model parameters.

\begin{table}[pt]
\begin{center}
{\begin{tabular}{@{}c|cc|ccc|c@{}} \hline
$\mu$\ / GeV & $M$ / GeV & $\Phi(M,U=0)$ & $M$ /GeV & $U$ /GeV & $\Phi(M,U)$  & $\Phi(M=0,U=0)$ \\
& ($M\neq 0,\ U=0$) & / GeV$^4$ & & & /GeV$^4$ & /GeV$^4$  \\ \hline
0.0\hphantom{000} & 0.322387 & $\underline{-0.0246944}$ & 0.279373 & 0.138605 & $-0.0246559$ & $-0.0240940$ \\
0.1\hphantom{000} & 0.322387 & $\underline{-0.0246944}$ & 0.279373 & 0.138605 & $-0.0246559$ & $-0.0240991$ \\
0.1408 & 0.322387 & $\underline{-0.0246944}$ & 0.279373 & 0.138605 & $-0.0246559$ & $-0.0241139$ \\
0.1409 & 0.322387 & $\underline{-0.0246944}$ & $-$ & $-$ & $-$ & $-0.0241140$ \\
0.1999 & 0.322387 & $\underline{-0.0246944}$ & $-$ & $-$ & $-$ & $-0.0241749$ \\
0.2\hphantom{000} & 0.322387 & $\underline{-0.0246944}$ & 0.286078 & 0.11997 & $-0.0246606$ & $-0.0241751$ \\
0.3\hphantom{000} & 0.322387 & $\underline{-0.0246944}$ & 0.311389 & 0.0431851 & $-0.0246902$ & $-0.0245044$ \\
0.32\hphantom{00} & 0.322387 & $\underline{-0.0246944}$ & 0.318850 & 0.0152899 & $-0.0246941$ & $-0.0246252$ \\
\hphantom{00}0.322387 & 0.322387 & $\underline{-0.0246944}$ & 0.320032 & 0.0103812 & $-0.0246943$ & $-0.0246412$ \\
0.3224 & $-$ & $-$ & 0.320039 & 0.0103560 & $\underline{-0.0246943}$ & $-0.0246413$ \\
0.323\hphantom{0} & $-$ & $-$ & 0.320391 & 0.00885381 & $\underline{-0.0246944}$ & $-0.0246454$ \\
0.324\hphantom{0} & $-$ & $-$ & 0.321140 & 0.00560727 & $\underline{-0.0246944}$ & $-0.0246523$ \\
0.3246 & $-$ & $-$ & 0.321754 & 0.00289703 & $\underline{-0.0246944}$ & $-0.0246564$ \\
0.3247 & $-$ & $-$ & $-$ & $-$ & $-$ & $\underline{-0.0246564}$ \\
0.4\hphantom{000} & $-$ & $-$ & $-$ & $-$ & $-$ & $\underline{-0.0253909}$ \\ \hline 
$\mu\ (>0.4)$ & $-$ & $-$ & $-$ & $-$ & $-$ & $-\frac{\mu^4+3\Lambda^4}{2\pi^2}$
\end{tabular}}
\caption{
The numerical results for the quark mass $M$, the pseudovector condensate $U$ and the thermodynamic potential $\Phi(M,U)$ are shown as a function 
of the quark chemical potential $\mu$. The underline for the numerical values of the thermodynamic potential means the lowest value of the thermodynamic potential 
in a few branch of the solutions. }
\end{center}
\end{table}

Next, let us assume $U\geq 0$. If the quark chemical potential $\mu$ is large, then, 
the dynamical quark mass becomes zero because the chiral symmetry is restored. 
Then, if $M=0$ and $\mu>U$ in the chiral limit, the thermodynamic potential with $B=0$ is written as 
\beq\label{5-3}
\Phi_{B=0}(M=0)
&=&
\frac{1}{2\pi^2}(-\mu^4+U^4+\Lambda^4+6\Lambda^2U^2)+\frac{1}{5\pi^2}\frac{1}{U}\left((\Lambda-U)^5-(\Lambda+U)^5\right)\nonumber\\
& &+\frac{U^2}{4G_p}\ . 
\eeq
Then, the gap equation for $U$ is obtained as 
\beq\label{5-4}
\frac{\partial \Phi_{B=0}(M=0)}{\partial U}=\frac{U}{5\pi^2}\left(2U^2-10\Lambda^2+\frac{5\pi^2}{2G_P}\right)=0
\eeq 
Thus, we obtain the solution 
\beq\label{5-5}
U=0\ , \qquad {\rm or}\qquad 
U=\sqrt{5\Lambda^2-\frac{5\pi^2}{4G_p}}\ . 
\eeq
In the following parts of this section, we adopt $G_p=2G_s$. 
Under these parameters with $\Lambda=0.631$ GeV and $G_s=5.5$ GeV$^{-2}$, 
a non-trivial solution of the gap equation gives 0.932 GeV for $U$. 
This value is larger than the cutoff $\Lambda$ and also the condition $\mu>U$ is not satisfied. 
Thus, if $M=0$, then only $U=0$ may be a possible solution.

\begin{table}[pt]
\begin{center}
{\begin{tabular}{@{}c|cccc@{}} \hline
$\mu$\ / GeV & $M$ / GeV & $U$ / GeV & $\Phi(M,U)$ /GeV$^4$ & ${\cal M}\times 10^{18}$ / (C/ms) \\ \hline
0.3224 & 0.320039 & 0.0103560 & $-0.0246943$ & 1.15849 \\
0.3226 & 0.320153 & 0.00987126 & $-0.0246943$ & 1.1046 \\
0.3228 & 0.320269 & 0.00937387 & $-0.0246944$ & 1.04871 \\
0.3230 & 0.320391 & 0.00885481 & $-0.0246944$ & 0.990414 \\
0.3232 & 0.320518 & 0.00830905 & $-0.0246944$ & 0.929145 \\
0.3234 & 0.320653 & 0.00772916 & $-0.0246944$ & 0.864079 \\
0.3236 & 0.320797 & 0.00710339 & $-0.0246944$ & 0.793903 \\
0.3238 & 0.320956 & 0.00641091 & $-0.0246944$ & 0.716291 \\
0.3240 & 0.321140 & 0.00560727 & $-0.0246944$ & 0.62679 \\
0.3242 & 0.321380 & 0.00454891 & $-0.0246944$ & 0.507837 \\
0.3244 & 0.321717 & 0.00305037 & $-0.0246944$ & 0.339844 \\
0.3246 & 0.321754 & 0.00289703 & $-0.0246944$ & 0.321587 \\ \hline
\end{tabular}}
\caption{
The numerical results for the quark mass $M$, the pseudovector condensate $U$, the thermodynamic potential $\Phi(M,U)$ and the spontaneous magnetization 
per unit volume ${\cal M}$ are shown as a function 
of the quark chemical potential $\mu$ in the range from $\mu=0.3224$ GeV to 0.3246 GeV, in which the phase with $M\neq 0$ and $U\neq 0$ 
is realized. 
}
\end{center}
\end{table}

%
\begin{figure}[b]
\begin{center}
\includegraphics[height=6.5cm]{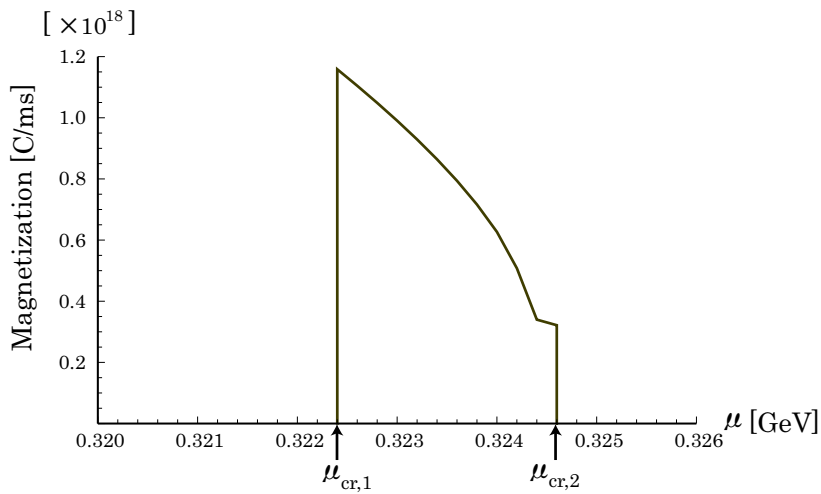}
\caption{The spontaneous magnetization per unit volume ${\cal M}$ is depicted as a function of the quark chemical potential $\mu$. 
}
\label{fig:fig1}
\end{center}
\end{figure}
%

The numerical results for the quark mass $M$, the pseudovector condensate $U$ and the thermodynamic potential $\Phi(M,U)$ are summarized in 
Table I as a function 
of the quark chemical potential $\mu$. 
The underline for the numerical values of the thermodynamic potential represent the lowest value of the thermodynamic potential 
in a few branch of the solutions and hyphen represents no solution. 
Usually, $\mu \leq 0.3224$ GeV $(=\mu_{\rm cr,1})$, the chiral symmetry is broken and the non-trivial solution of the gap equation for 
chiral condensate or dynamical quark mass exists. 
Also, in $\mu > \mu_{\rm cr,1}$, the chiral symmetry is restored and only $M=0$ has a true solution. 
However, in the case with the pseudovector interaction, other solutions exist. 
Namely, in larger region of the quark chemical potential $\mu > \mu_{\rm cr,1}$, 
the chiral symmetry is still not restored and the solution of the gap equation for the dynamical quark mass with $M\neq 0$ appears 
with $U\neq 0$. 
The window of the quark chemical potential where the non-trivial solutions with $M\neq 0$ and $U\neq 0$ exist is very narrow.  
In the region with $\mu > 0.3247$ GeV $(=\mu_{\rm cr,2})$, only the trivial solution with $M=U=0$ exists, which corresponds to the free quark gas. 
It should be noted here that the critical baryon density from the pseudovector condensed phase or spin polarized phase with $M\neq 0$ and $U\neq 0$ 
to the chiral symmetric phase with $M=U=0$ corresponds to $1.78 \rho_0$ where $\rho_0=0.17$ fm$^{-3}$ is the normal nuclear density. 
However, the two critical chemical potentials, $\mu_{{\rm cr}1}$ and $\mu_{{\rm cr},2}$ correspond to rather small quark number densities, $0.0116$ fm$^{-3}$ and
$0.0030$ fm$^{-3}$, respectively. 
According to Ref. \citen{Costa} where the first order phase transition was studied in the $\rho$-$T$ and 
$\mu$-$T$ space, these small densities lie inside the low density metastable before the onset of the spinodal region. 

In this narrow window, the spontaneous magnetization per unit volume (\ref{4-13}) with $\Phi$ in (\ref{4-14}) instead of $\Phi_A$ appear. 
The numerical results are summarized in Table II. 
The order of magnitude of the spontaneous magnetization is about $10^{17}$ and/or $10^{18}$ C/ms. 
These magnitude leads to the magnetic flux density with $10^{13}$ or $10^{14}$ Gauss in the surface of compact stars\cite{oursPTEP3}.
Also, the result of the spontaneous magnetization per unit volume is shown in Fig.1. 
It should be noted here that, if the coupling strength $G_p$ is smaller than the value adopted here, the window, in which the spontaneous 
magnetization occurs, does not open. 
Namely, there is no pseudovector condensate or spin polarized condensate.
On the other hand, if $G_p$ is rather large, the local minimum of the thermodynamic potential with respect to finite $U$ and $M$ changes to 
the saddle point as $G_p$ increases. 
Thus, the pseudovector condensate only exists in very narrow region in the parameter space, $G_p$.

\subsection{Magnetic susceptibility}

We calculate the magnetic susceptibility in the same way as spontaneous magnetization.
First, we define the magnetic susceptibility as
\begin{align}
	\chi &= \ \mu_0\left.\frac{\partial \mathcal{M}}{\partial B}\right|_{B=0} =-\mu_0\left.\frac{\partial^2\Phi_A}{\partial B^2}\right|_{B=0},
\end{align}
where $\mu_0$ represents the vacuum permeability. 
From (\ref{21}), specifically, it is obtained as follows: 
\begin{eqnarray}
\Delta\chi &\equiv& \chi-\chi_{(\mu=0)}\nonumber\\
&= &-\frac{5e^2\mu_0}{36\pi^2}\left[\frac{\sqrt{(\mu-U)^2-M_q^2}}{\mu}-\int^{\sqrt{(\mu-U)^2-M_q^2}}_0 dp_z \frac{1}{\sqrt{p_z^2+M_q^2}+U} \right]
\nonumber\\
& &
\qquad\qquad\qquad\qquad\qquad\qquad\qquad\qquad\qquad\qquad\qquad\qquad
\times
 \theta\left(\mu-(M_q+U)\right) \nonumber \\
& &-\frac{5e^2\mu_0}{36\pi^2}\left[\frac{\sqrt{(\mu+U)^2-M_q^2}}{\mu}-\int^{\sqrt{(\mu+U)^2-M_q^2}}_0 dp_z \frac{1}{\sqrt{p_z^2+M_q^2}-U} \right] 
\nonumber\\
& &
\qquad\qquad\qquad\qquad\qquad\qquad\qquad\qquad\qquad\qquad\qquad\qquad
\times
\theta\left(\mu-(M_q-U)\right) \nonumber \\
& &+\frac{5e^2\mu_0}{36\pi^2}\int^\Lambda_0 dp_z \Biggl[ \left( \Lambda^2 - p_z^2 + \left( \sqrt{p_z^2+M_q^2} + U\right)^2\right)^{-\frac{1}{2}} 
\nonumber\\
& &\qquad\qquad\qquad\qquad
+ \left( \Lambda^2 - p_z^2 + \left( \sqrt{p_z^2+M_q^2} - U\right)^2\right)^{-\frac{1}{2}} \Biggl] \nonumber \\
	& &-\frac{5e^2\mu_0}{36\pi^2}\int^\Lambda_0 dp_z \left[\frac{1}{\sqrt{p_z^2+M_q^2}+U}+\frac{1}{\sqrt{p_z^2+M_q^2}-U}\right] - \chi_{(\mu=0)}.
\end{eqnarray}
\begin{table}[pt]
\begin{center}
{\begin{tabular}{@{}c|cccc@{}} \hline
\ \ \ \ $\mu$\ / GeV & $M$ / GeV & $U$ / GeV & $\Delta\chi \times 10^{-6}$  \ \ \\ \hline
0.3222 & 0.322387 & 0 &  0 \\
0.3224 & 0.320039 & 0.0103560 &  $-5.80492$ \\
0.3226 & 0.320153 & 0.00987126 & $-5.42174$ \\
0.3228 & 0.320269 & 0.00937387 & $-5.03584$ \\
0.3230 & 0.320391 & 0.00885481 & $-4.64583$ \\
0.3232 & 0.320518 & 0.00830905 & $-4.24974$ \\
0.3234 & 0.320653 & 0.00772916 & $-3.84460$ \\
0.3236 & 0.320797 & 0.00710339 & $-3.42554$ \\
0.3238 & 0.320956 & 0.00641091 & $-2.98378$ \\
0.3240 & 0.321140 & 0.00560727 & $-2.50005$ \\
0.3242 & 0.321380 & 0.00454891 & $-1.91054$ \\
0.3244 & 0.321717 & 0.00305037 & $-1.09987$ \\
0.3246 & 0.321754 & 0.00289703 & $-0.60179$ \\
0.3248 & 0 & 0 & $-336.097$  \\ \hline
\end{tabular}}
\caption{
The numerical results for the quark mass $M$, the pseudovector condensate $U$
and the magnetic susceptibility 
${\Delta\chi}$ by the positive energy quarks are shown as a function 
of the quark chemical potential $\mu$ in the range from $\mu=0.3222$ GeV to 0.3248 GeV. }
\end{center}
\end{table}
%
\begin{figure}[b]
\begin{center}
\includegraphics[height=5.5cm]{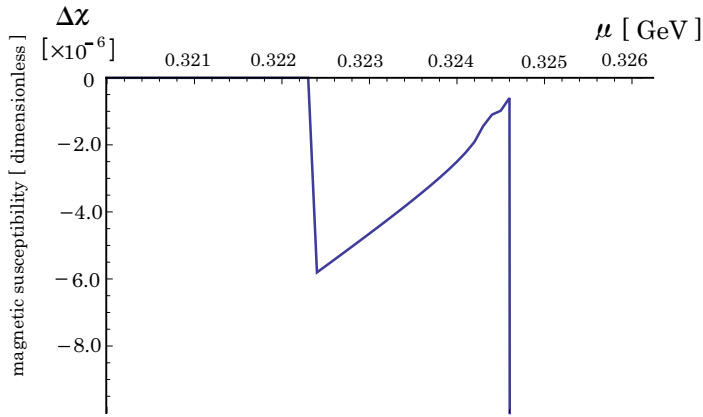}
\caption{The magnetic susceptibility compared with vacuum value $\chi_{(\mu=0)}$, $\Delta\chi$, is depicted as a function of the quark chemical potential $\mu$. 
}
\label{fig:fig2}
\end{center}
\end{figure}
%
Here we subtract the following $\chi_{(\mu=0)}$ in order to investigate the contribution only of the positive-energy particles due to the pesudvector-type interaction between quarks:
\begin{align}
	\chi_{(\mu=0)} = 
-\frac{5e^2\mu_0}{18\pi^2}\left[-\frac{\Lambda}{\sqrt{\Lambda^2+M_q^2(\mu=0)}}+ \ln\frac{\Lambda+\sqrt{M_q^2(\mu=0)+\Lambda^2}}{M_q(\mu=0)}\right]\ .
\end{align}
In the region with $\mu<\mu_{\rm cr,1}$, the solutions of gap equations have the values $M\neq0$ and $U=0$, so $\chi_{\mu<\mu_{\rm cr,1}} \equiv \chi_<$ is obtained as
\begin{align}
	\Delta\chi_{<} &= -\frac{5e^2\mu_0}{18\pi^2}\left[-\frac{\Lambda}{\sqrt{\Lambda^2+M_q^2}}+ \ln\frac{\Lambda+\sqrt{M_q^2+\Lambda^2}}{M_q}\right] - \chi_{(\mu=0)} \nonumber \\
	&=0\ 
\end{align}
due to a subtraction of the contribution of the vacuum $\chi_{(\mu=0)}$. 
In the same way, in the region with $\mu> \mu_{\rm cr,2}$, the solutions with $M=U=0$ exist. Then $\chi_{\mu> \mu_{\rm cr,2}} \equiv \chi_>$ is  obtained as
\begin{align}
	\Delta\chi_{>} = -\frac{5e^2\mu_0}{18\pi^2}\ln\frac{\Lambda}{\mu} - \chi_{(\mu=0)}.
\end{align}
The numerical results are summarized in Table III. The order of magnitude is about $10^{-6}$ in the region with $\mu_{\rm cr,1} \lesssim \mu \lesssim \mu_{\rm cr,2}$. 
Also, the result of the magnetic susceptibility is shown in Fig.\ref{fig:fig2}.

\section{Summary and concluding remarks}

It has been shown that the spontaneous magnetization occurs due to the pseudovector-type four-point interaction between quarks 
in quark matter at zero temperature within the NJL model. 
In the narrow region of the quark chemical potential, both the chiral condensate and pseudovector condensate, namely spin polarized 
condensate, coexist, which leads to the spontaneous magnetization. 
On the contrary, in the tensor-type four-point interaction between quarks, the spin polarization occurs above a
certain quark chemical potential, that is in the high density quark matter. 
However, the spontaneous magnetization does not appear in the case of the tensor interaction 
except for the existence of the anomalous magnetic moment of quarks \cite{oursPTEP3}. 
Also, we have calculated the magnetic susceptibility by expanding the thermodynamic potential up to the second order of the external magnetic field. 
As a result, the Landau diamagnetism may be revealed because only the contribution of the positive-energy quarks being free quasi-particles is considered. 

In this paper, we ignore the effects of current quark mass. 
It was pointed out that the region in which the pseudovector condensate has non-zero value enlarges, if the current quark mass is introduced \cite{Maedan}. 
Further, the effects of the strange quark is missing in this work. 
These are interesting future problems which are left in order to clarify the magnetic properties of high density quark matter.

\section*{Acknowledgements}

One of the authors (Y.T.) would like to express their sincere thanks to\break
Professor J. da Provid\^encia and Professor C. Provid\^encia, two of co-authors of this paper, 
for their warm hospitality during their visit to Coimbra in spring of 2016. 

\appendix



\end{document}